\documentclass[epj]{svjour}
\usepackage{graphicx}
\usepackage{graphics}
\usepackage{amsmath}
\usepackage{amssymb}
\usepackage{latexsym}
\usepackage{amsbsy}
\usepackage{amsfonts}
\begin{document}
\title{Pairing Correlations in Finite Systems: From the weak to the strong fluctuations regime}
\author{M.A. Fern\'andez and J.L. Egido}
\institute{Departamento de F\'{\i}sica Te\'orica C-XI, Universidad 
Aut\'onoma de Madrid, 28049 Madrid, Spain} 
\authorrunning{M.A. Fern\'andez and J.L. Egido}
\titlerunning{Pairing Correlations in Finite Systems}
\abstract{
The Particle Number  Projected Generator Coordinate Method is formulated
for the pairing Hamiltonian in a detailed way  in the projection after variation 
 and the variation after projection methods.  
  The dependence of the wave functions  on the generator coordinate is analyzed  performing 
 numerical applications for the most relevant collective  coordinates. The calculations reproduce 
 the exact solution  in the weak, crossover and strong pairing regimes.
The physical insight of the Ansatz and its numerical simplicity make this theory 
an excellent tool to study pairing correlations in  complex situations and/or involved 
Hamiltonians.
}
\maketitle
%
\section{Introduction}

The measurements of Black, Ralph and Tinkham \cite{RBT1,RBT2}   of discrete level spectra and spectroscopic gaps
in  nanometer Al isolated grains were interpreted as  evidence  of the superconductivity phenomenon. To understand
 the physics of such ultrasmall grains a great deal of  theoretical effort was devoted  to study such systems 
 starting from grand canonical (BCS) and canonical ensembles \cite{MFF.98}  
as well as  very sophisticated theories \cite{DS.83} of the pairing Hamiltonian, see ref.~ \cite{vD.01} 
for a review. Later on the exact solution \cite{RS.64} of this  naive model Hamiltonian  was rediscovered.
    Some others studies treat the aspect of thermodynamic properties \cite{SIL.63,FFG.00}  
while  others, see for example  \cite{RCE.01},  the question of persistence of
 pairing correlations above the BCS critical temperature is addressed.   More recently some analytical results
  in special regimes have been obtained  \cite{YBA.05}. 
  The main issue of all these studies is the proper description of the crossover between the few electron 
  regime and the bulk one. To analyse this crossover several
properties can be computed as a function of the mean electronic level spacing $d$  (or the number of
 electrons $N$) that characterizes the transition from one 
regime to the other.  One of the  findings of these studies was that the strong phase transition predicted 
in  a grand canonical study was absent in more 
advanced theories as well as  in the exact solution.  In the BCS approach superconductivity is not  possible
 for all $d$ ($N$)  breaking down at a  critical $d$ value. 
This break down is number parity dependent and  indicates  that quantum fluctuations are  not treated 
adequately by the BCS wave function.  The knowledge of the exact solution
for the simple-minded pairing Hamiltonian does not diminish importance to the  theoretical approximations
 developed for the study of that Hamiltonian, see \cite{BD.99a} for a review.  These approximations are very 
 general and allow  the study of more sophisticated Hamiltonians for which no exact solution exists. The use of
exactly solvable Hamiltonians\cite{DES.01}, on the other hand, is very practical since it allows to check the
 accuracy of different approximations in the limiting cases represented by those Hamiltonians.
 
 In a recent  paper \cite{FE.03}  we  have proposed a new approach to study superconductivity in finite
 systems,  namely  the 
Generator Coordinate Method (GCM) \cite{Hiwhe}, 
based on particle number projected BCS wave functions generated in a suitable way. In that paper the GCM approach
was applied to superconducting grains described by  the Pairing Hamiltonian and  it was  shown to  provide   
an accurate description of these systems in  perfect  agreement with the exact Richardson solution.
The purpose of this paper is two-fold, first, to present a detailed derivation of the relevant formula as well 
as the way to solve the 
Hill-Wheeler (HW) equations and, second, to analyse different generator coordinates in the context of  pairing 
correlations.   The derivation
presented is comprehensive  enough to allow for the application of the formalism to  other pairing Hamiltonians. 
Furthermore since our theory is very general and not constrained by any  requirement  can be applied to more
 complex systems. As a matter of fact we have performed preliminary
studies with the most general pairing Hamiltonians proposed in  \cite{DES.01} and the results \cite{FE.06} are of 
the same quality as the ones presented in this investigation. 
Finite temperature effects are not considered in the present study.

In sect.~\ref{sect:theory} we derive the general formula of the GCM. In sect.~\ref{sect:sel}
 we discuss the different coordinates to be used in the calculations.   The convergence
 and other issues concerning the numerical solution of the HW equations is analysed in sect.~\ref{sect:num}.
  Finally in sect.~\ref{sect:app} the whole formalism is applied to study superconducting grains. The paper ends
  with the Conclusions and some numerical aspects discussed in the Appendices A and B.

\section{Theory}
\label{sect:theory}
The pairing Hamiltonian used in most calculations is given by 
\begin{equation}\label{PH}
H = \sum_{k=1,\nu=\pm}^{N} \epsilon_k c^\dagger_{k,\nu} \; c_{k,\nu}
- G \sum_{k,k'=1}^{N} \; c^\dagger_{k+} c^\dagger_{{k-}} c_{{k'-}}
c_{k'+}
\end{equation}
where $k+$ (${k-}$) labels the single particle level (time reversed)  with energies $\epsilon_k$ and $c_{k}, c^\dagger_k$
 destroys and creates electrons in their respective states. The    interaction constant 
$G$ is taken as $\lambda \, d$ with $d$ the level spacing and $\lambda$ the BCS coupling constant whose value for Al is
$0.224$. The single particle energies $\epsilon_k$
for simplicity  take the values $\epsilon_k=kd$. The number $N$  of electrons is equal to the number of levels and in 
the ground state  they form $N/2$ Cooper pairs, so one works at half filling.  This Hamiltonian allows the discussion 
of the crossover between  the strong-coupling regime $(d/{\tilde{\Delta}} \ll 1)$ that represents  large grains and the 
weak-coupling regime $(d/{\tilde{\Delta}} \gg 1)$ for small grains,  in terms of the quantity 
$d/\tilde{\Delta}=2 {\rm sinh}(1/\lambda)/N$  with $\tilde{\Delta}$ the bulk gap, or equivalently in terms of the number
 of electrons $N$.

The simplest way to deal with pairing correlations  is provided by the BCS theory \cite{BCSA}.  Its Ansatz is given by the
 mean field wave function
\begin{equation}
\label{BCS}
| BCS \rangle_{\varphi} = \prod_{k > 0}^{N} (u_k + v_k e^{i\varphi}c_{k+}^{\dagger} 
c_{k-}^{\dagger}) |-\rangle.
\end{equation}   
The variational parameters $v_k$ are related to the  probability to find two electrons in the level $k$. The parameters 
$u_k$ are given by $u_k^2 + v_k^2 =1$. The spontaneous particle number symmetry breaking mechanism implicit in Eq.~\ref{BCS} 
 enlarges the available variational Hilbert space  making the BCS approximation, in the case of large particle numbers, a 
 very good one.  The BCS state (\ref{BCS}), on the other hand, undergoes strong 
particle number fluctuations and for finite systems like metal grains, the  Ansatz (\ref{BCS}) is unreliable and  misses
 essential features.  To correct this failure it is necessary to develop the BCS formalism in a canonical ensemble,  where 
 the particle number is fixed, rather than in a grand-canonical one.   The restoration of the particle number in the BCS
  context was introduced  by Dietrich and Mang \cite{DM.64} in a nuclear structure context  and it was applied for the first 
  time to superconducting grains by J. von Delft and F. Braun \cite{BD.98,BD.99b}.  The  projection method is based  on the 
  Anderson formulation of superconductivity \cite{ANDERSON58} where projection onto good particle number is presented 
  as an integration in the gauge variable $\varphi$,
\begin{equation}
| BCS \rangle_N =  \!\! \int_0^{2\pi} \frac{d \varphi}{2\pi}  \, e^{iN\varphi} 
\prod_k^N (e^{-i\varphi/2} u_k +
e^{i\varphi/2}v_k c_{k+}^{\dagger} c_{k-}^{\dagger})|-\rangle
\label{eq:BCSN}
\end{equation}      
We assume the number of particles $N$ to be even, the odd case  is considered in appendix \ref{App:PBCSOdd}.  
The formulation of the particle number projection (PNP) can be done in several ways\cite{RS.80}. Very compact
formula are obtained  in terms of the \emph{residuum integrals} \cite{DM.64} defined by
\begin{eqnarray}
\label{resint}
\nonumber
R_m^{j_{1}, \cdots, j_{M}} 
&=& \frac{1}{2 \pi} \int_{0}^{2 \pi} d \varphi \, 
e^{-i(M - 2m)\varphi/2}  \\ 
&\times& \prod_{k \neq j_1, \cdots, j_M}^N (e^{-i\varphi/2}u_k^2 
+ e^{i\varphi/2} v_k^2)       
\end{eqnarray}
This definition holds for indices ${j_{1}, \cdots, j_{M}}$ such that
$j_k \neq j_p$ for all $k$ and $p$.  The integer $M$ is simply a  counter of the $j$'s involved. In case that two or more indices are
equal we {\it define} the corresponding residuum integral as zero.
All expectation values can be easily calculated  in terms of the residuum 
integrals.  As an example we evaluate the matrix element \(_N\langle BCS | BCS \rangle_N \), direct substitution of Eq.~\ref{eq:BCSN}
provides
\begin{equation}
_N\langle BCS | BCS \rangle_N =   \int_{0}^{2 \pi} \frac{d \varphi}{2 \pi} \, \prod_{k}^N (e^{-i\varphi/2}u_k^2 
+ e^{i\varphi/2} v_k^2)\equiv R_{0}^{0}.
\label{NNorm}
\end{equation}    
In the same way the projected energy is given by 
\begin{eqnarray}
\nonumber
E_N &=& \frac{_N\langle BCS |H| BCS \rangle_N}{_N\langle BCS | BCS \rangle_N} \\ 
    &=&  2 \sum_{j=1}^N \left( \epsilon_j - \frac{G}{2} \right) v_j^2 \frac{R_1^j}{R_0^0} -
G \sum_{j,k}^N u_j v_j u_k  v_k \frac{R_1^{jk}}{R_0^0}.
\label{Eproj}
\end{eqnarray}

The PNP energy, as the BCS one, depends only on the variational parameters
 $u_k, v_k$.  Minimization with respect to these parameters leads  to a set of
$N$ coupled non-linear equations 
\begin{equation}
\label{VAR_EQU}
2(\hat{\epsilon}_k + \Lambda_k)u_k v_k - \Delta_k 
(u_k^2 - v_k^2)  = 0.
\end{equation}
The quantities $\hat{\epsilon}_k$, $\Delta_k$ and $\Lambda_k$
 are defined by 
\begin{equation}
\hat{\epsilon}_k =  (\epsilon_k - G/2)\frac{R_1^k}{R_0^0}, \,\,\,\,\,\,\,\,\,\,
\Delta_k = G \sum_{j} u_j v_j \frac{R_1^{kj}}{R_0^0}
\end{equation}

\begin{eqnarray}
\Lambda_k &=& \sum_j^N \left(\epsilon_j - \frac{G}{2}\right)v_j^2 \left[ 
\frac{R_2^{kj}-R_1^{kj}}{R^0_0} - \frac{R_1^{j}}{R^0_0} \frac{R_1^k - R_0^k}
{R_0^0} \right] \nonumber \\
 &-&  \frac{G}{2} \sum_{j,l}^N u_j v_j u_l v_l \left[
\frac{R_2^{kjl} - R_1^{kjl}}{R_0^0} - \frac{R_1^{jl}}{R_0^0} 
\frac{R_1^k - R_0^k}{R_0^0}\right] 
\end{eqnarray}

The set of equations (\ref{VAR_EQU})  resembles the ordinary BCS equations. In that equations, $\Lambda_k=0 $ and 
$\hat{\epsilon}_k = \epsilon - G v_k^2 - \mu$. The Lagrange multiplier $\mu$ 
takes care, on the average, of the particle number conservation.
Notice that in the projected equations the fields $\Lambda_k$ appear  in addition.
The solution of Eqs.(\ref{VAR_EQU}) defines $| BCS \rangle_N$.  In the literature \cite{BD.98} this is usually called Projected BCS (PBCS) theory.
Details of how the set of Eqs.(\ref{VAR_EQU}) is 
numerically solved   are given in appendix \ref{App:NumSol}. 

To   include additional   correlations  we consider a \emph{general superposition} of different projected-BCS wave functions,
\begin{eqnarray}
\nonumber
| \Psi_N \rangle  & =&   \int d\xi   \,\, f(\xi)\,\, | BCS (\xi) \rangle_N \label{ansatz} \\  
\nonumber
& =&  \frac{1}{2\pi}\,\int  d\xi \,\, d\varphi \,\,  f(\xi) \,\, 
e^{iN\varphi} \\
&\times& \prod_k (e^{-i\varphi/2} u_k(\xi) + e^{i\varphi/2} v_k(\xi) c_{k+}^\dagger  c_{k-}^{\dagger}) |-\rangle. 
\label{BCSANSA}
\end{eqnarray}

The new wave function $| \Psi_N \rangle$ is based on the Generator Coordinate Method 
(GCM) developed by Hill and Wheeler in Nuclear  Physics \cite{Hiwhe}.  
It has been also used by Peierls, Yoccoz and Thouless \cite{PY.57,PT.62} among others
to deal with the restoration of symmetries in mean field approaches  as
 well as to deal with a variational approach to collective motion. It has also 
provided a variational derivation of the Random Phase Approximation\cite{JS.64}.
The coordinate $\xi$ refers to any parameter
on which the BCS states may depend parametrically. In this way  the superposition state $| \Psi_N \rangle$ takes care of  the 
fluctuations associated to the parameter $\xi$. In principle, the variational quantities are the weights $f(\xi)$
and the occupancies $u_k(\xi), v_k(\xi)$ and should be determined invoking the variational principle. 
The final equations, however, result in an integro-differential set of equations very complicated to solve.
In consequence some assumptions about occupancies are needed in order to facilitate the numerical 
implementation.  If we assume that the quantities  $u_k(\xi), v_k(\xi)$ are known (see below) one deals only
with the problem of calculating the weights  $f(\xi)$.  This is accomplished by the Hill-Wheeler (HW) equation
\begin{equation}
\label{HWEQ}
\int d\xi \, \,(\mathcal{H}_{\xi \xi^{'}}
 - \mathcal{E} \mathcal{N}_{\xi
\xi^{'}}) f(\xi) = 0.
\end{equation} 

$\mathcal{H}$ is the Hamiltonian  overlap, defined by
\begin{eqnarray}
\mathcal{H}_{\xi \xi^{'}} \!
 & = & \!  _N\langle BCS(\xi) |H| BCS (\xi')\rangle_N \nonumber \\
 & = & 2 \sum_j (\epsilon_j - \frac{G}{2})v_j(\xi)v_j(\xi^{'})
 R_1^j(\xi,\xi^{'}) \nonumber \\  
&-& G \sum_{i,j,i\ne j} u_i(\xi^{'}) v_i(\xi) u_j(\xi) v_j(\xi^{'})
R_1^{ij}(\xi,\xi^{'})
\label{hamove}
\end{eqnarray}
and $\mathcal{N}$ the norm overlap
\begin{equation}
\mathcal{N}_{\xi \xi^{'}} \!  =  \! _N\langle BCS (\xi) |BCS (\xi')\rangle_N 
\! = \! {R_0^0 (\xi,\xi^{'})}.
\label{normove}
\end{equation}

This equation  is very similar to  Eq.~(\ref{NNorm}). The residuum integrals have  now been generalized by

\begin{eqnarray}
\label{genresint}
 &&R_m^{j_{1}, \cdots, j_{M}} (\xi,\xi^{'}) 
= \frac{1}{2 \pi} \int_{0}^{2 \pi} d \varphi \, 
e^{-i(M - 2m)\varphi/2} \nonumber \\
&\times& \!\!\!\!\!\!\!\! 
\prod_{k \neq j_1, \cdots, j_M} \!\!\! (e^{-i\varphi/2}u_k(\xi)u_k(\xi^{'}) 
+ e^{i\varphi/2} v_k(\xi)v_k(\xi^{'}))       
\end{eqnarray}
In App.~\ref{app:HamOver} we discuss the way to calculate these integrals
and the Hamiltonian overlap of Eq.~(\ref{hamove}). It is important to notice
that the solution of the HW equation provides not only the ground state but
also the low-lying collective states.

As we mentioned above,  in the HW equation the quantities 
 $u_k(\xi), v_k(\xi)$ are supposed to be determined beforehand.
They are usually fixed by the way  the  projected wave function $|BCS(\xi)\rangle_N$ is calculated,
namely,  whether  $|BCS(\xi)\rangle_N$ is determined by \emph{projection after 
variation} (PAV) or \emph{variation after projection} (VAP).  In the former
 (PAV), the occupancies are determined by the symmetry-violating wave function
$|BCS\rangle$, i.e., by solving the ordinary BCS equations. In the VAP 
case the occupancies are given by the solution of  the variational equations 
Eq.(\ref{VAR_EQU}).  In the BCS framework the VAP approach is known as PBCS.
  Obviously the VAP method is more involved but it is a fully self-consistent method that provides better results.  We shall denote the first 
method GCMPAV  and the second one  GCMVAP.

\section{SELECTION OF THE GENERATOR COORDINATE }
\label{sect:sel}
The generator coordinate $\xi$ is quite general and its  selection is motivated by the physical problem. The BCS
wave functions depend parametrically on the generator coordinate, its selection is therefore strongly related to the ways we have to characterize
the wave function. Though there are many ways to choose the generator coordinate,
we think that for the BCS case there are three  relevant ones:  The gap parameter $\Delta$, 
the Lagrange parameter $\mu$ associated with the particle number of the BCS wave function and 
 $\Delta N^2= \langle BCS |\hat{N}^2| BCS \rangle-  \langle BCS |\hat{N}| BCS \rangle^2$, the fluctuations on the
number of particles of the BCS wave function.

Instead of using directly the gap parameter as a coordinate it is numerically easier to generate BCS w.f.'s with 
different gap parameters by solving the corresponding BCS (PBCS) equations  for different values $G_{trial}$  of the strength constant $G$. This
method is easy to implement and very efficient. The second method is the simplest one.  Now the generator
coordinate is the chemical potential $\mu$ which in the ordinary BCS equations is used  as
Lagrange multiplier to fix the mean value of the particle number in the grand-canonical ensemble. In our case we solve 
the BCS equations {\em for fixed} $\mu$ and the use of different $\mu$ values allows to generate wave functions $|BCS(\mu)\rangle$ with
different average particle number. The fact that $|BCS(\mu)\rangle$ does not have on the average the right particle number does not matter
since later on we project  on the right particle number.  In this case one is
looking for the fluctuations in the position of the Fermi level.  
 
\begin{figure}[]
\begin{center}
\includegraphics[scale=0.6]{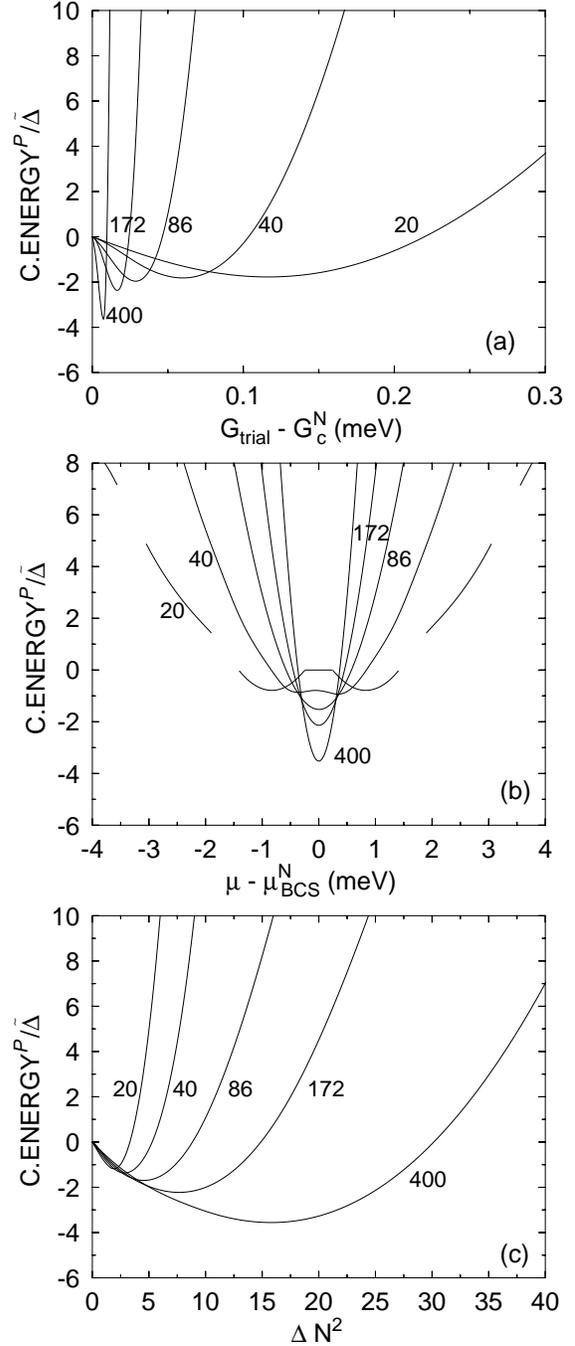}
\caption{Projected condensation energies, in units of the bulk gap,  as functions of the different generator coordinates in the PAV approach.}
\label{fig1}
\end{center}
\end{figure}

 The last method, finally,  considers fluctuations around
the uncertainty in the particle number  $\Delta N^2$.
Now it is necessary to add a constraint   to fix a given value of $\Delta N^2$.  This is
done by using the modified Hamiltonian $H' = H - \mu N - \mu_2 \Delta N^2$,  where
the parameter $\mu_2$ guaranties  that the  constraint is fulfilled.
In principle we have set up six variational methods (PAV and VAP versions of each coordinate) but only five 
are feasible, because the VAP version of  $\mu$ (by construction) is not possible.    
 Although the ultimate test of the quality of the selection of the generator
coordinate will be the eigenstates of the HW equation it is interesting to
have a look on the Hilbert space generated by the different coordinates. 
The diagonal  elements of the matrix 
$\mathcal{H}_{\xi \xi^{}}/\mathcal{N}_{\xi \xi^{}}$ of the
HW matrix, Eq.~(\ref{HWEQ}), are the projected total  energies ${E}_N(\xi)$. 
This quantity is related to the condensation energy (CE) by 
${E}_{con}(\xi) = E_N(\xi) -E_{F} $
 with $E_{F}$  the uncorrelated energy of the Fermi sea, i.e., 
 $E_{F} = 2 \sum_j \epsilon_j - G N/2$.
  Though, as mentioned above, we have five ways to generate
w.f. depending parametrically  on $\xi$ we shall concentrate in this section on
the three PAV cases corresponding to the three different coordinates under study.
In Fig.\ref{fig1} we display ${E}_{con}(\xi)$ as a function of the 
corresponding generator coordinate $\xi$  and for different particle (level) numbers
to cover the full range from weak to strong pairing regimes.  For
simplicity we plot only the curves for grains with an even number of particles. 
 Let us first discuss the coordinate  $G_{trial}$. 
 It is obvious that, for each  number of particles $N$, a critical  value $G_{c}(N)$
of $G_{trial}$  exists  such that no superconducting solution of the system is 
found below it. In panel a) we show the CE versus
$G_{trial}-G_{c}(N)$. We find a parabolic behavior with the
vertex moving to  larger values of  $G_{trial}-G_{c}(N)$
as the particle number decreases (as one would expect). The curves get softer 
with decreasing particle number with the curve $N=20$  being  specially soft.
We also find that the value of the CE in the minimum is larger (in absolute value) as the particle number
increases (as one also would  expect).

 In panel b) the quantity ${E}_{con}(\mu)$ is plotted against $\mu-\mu_{BCS}(N)$,  $\mu_{BCS}(N)$ being the 
chemical potential of the BCS equation for the corresponding case.  Because of the particle-hole symmetry of the model the  subtraction of $\mu_{BCS}(N)$ provides  symmetric curves around $\mu-\mu_{BCS}(N)=0$ \footnote{It is important to notice that in this  case the numbers 20, 40, etc correspond {\em only}
to the number of levels and {\em not} to the number of particles of  $|BCS (\mu)\rangle$. Since we work 
without constraint on the particle number in general   $\langle BCS (\mu)|\hat{N} |BCS (\mu)\rangle \ne N $. On the other hand since
we are projecting on the particle number the wave functions $|BCS (\mu)\rangle_N$  correspond to a system with N particles.}.
For large $N$ the solutions are  parabola like curves which soften with decreasing particle number.
For $N\ge 40$ we find superconducting solutions for all $\mu$ values.  This is not the case for $N=20$ where for certain
$\mu$ intervals  we do not obtain any solution for the BCS equation, see below  for more details. This is not surprising because the standard 
selfconsistent
BCS equation does not provide  a correlated solution in this case, see below. 
Lastly in panel c)  ${E}_{con}(\Delta N^2)$ is plotted against 
$\Delta N^2$. Here we also obtain a parabolic behavior similar to the case a) 
with the difference that the minima shifted to large $\Delta N^2$ correspond
to the large particle numbers. The CE gets softer with larger particle
number as one would expect. 

 It is clear that the energy minima of the different coordinates provide an approximation to an unconstrained VAP calculation.
In Table~\ref{table1} we have summarized the minima of the parabola as well as  the VAP values and the exact ones.  We find that all three 
coordinates do
a good job for large particle numbers and that  big differences appear for small particle numbers, i.e., in the weakly correlated regime.   
 We find that in general and at this level the coordinate $G_{trial}$ is the most effective  followed by $\Delta N^2$ and $\mu$. 
Of course this does not mean very much since the configuration mixing calculations will change these results.

\begin{table}
\begin{center}
\caption{Condensation energies, in units of $\tilde{\Delta}$,  predicted by PAV, VAP and exact calculations.}
{\begin{tabular}{|c|c|c|c|c|c|c|c|c|c|} 
\hline
$N$ & $20$  & $40$ & $86$ &
$172$ & $400$ \\ \hline
$G_{trial} $ & -1.7716 & -1.8194 & -1.9053 & -2.3566 & -3.4192 
\\ 
$\mu$ & -0.7864 &  -0.9272 & -1.4925 & -2.0392 & -3.4625 
\\
$\Delta N^2$ & -1.1438 & -1.3654 & -1.6906 & -2.2227 & -3.5564 
\\
VAP        & -2.0625 & -2.2441 & -2.4015 & -2.5428 & -3.6551
\\
exact        & -2.2026 & -2.5284 & -2.9403 & -3.5322 & -4.8891
\\
\hline
\end{tabular}}
\label{table1}
\end{center}
\end{table}

\begin{figure}[tbp]
\begin{center}
\includegraphics[scale=0.6]{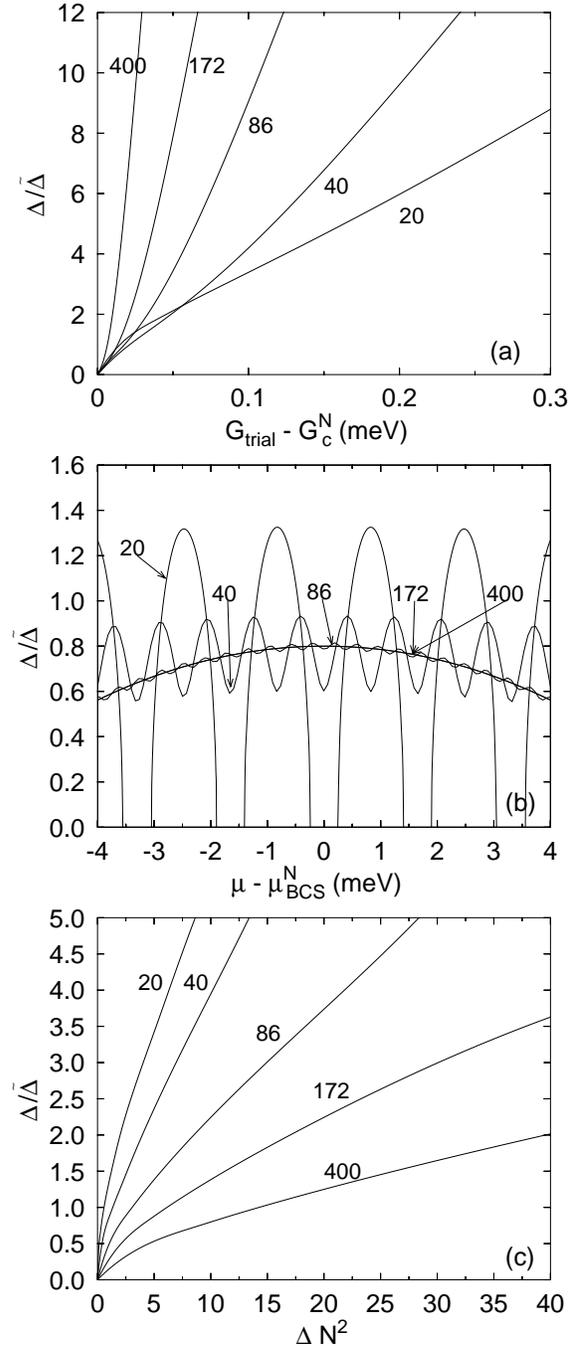}
\caption{Dependence of the order parameter $\Delta$ on
the generator coordinates $G_{trial}$ , $\mu$ and  $\Delta N^2$.  In panel (b) the y-axis scale applies only
for $N=20$, for $N=40$ the y-axis covers the interval $0.6-1.4$ and for $N=86, 172$ and 400 the interval $0.9-1.1$.
}
\label{fig2}
\end{center}
\end{figure}	

Let's now analyse  the wave functions generated with the different coordinates. 
To a given value of a coordinate, let say $\xi_0$, corresponds a wave function $|BCS(\xi_0>$.
A simple way to characterize the physical content of this wave function is by the associated gap
parameter  $\Delta(\xi_0) = G \sum_k u_k(\xi_0) v_k(\xi_0)$. In Fig.~\ref{fig2} we have represented 
the gap parameter  $\Delta(\xi)$ associated to each wave function as a function of the 
coordinate $\xi$ used to generate it. In panel a)
we show the results for $G_{trial}$.  Of course the $G$ entering into $\Delta$
is the one of the original Hamiltonian, see Eq.~({\ref{PH}), independently of the
$G_{trial}$ used in the calculations. 
Taking into account the expression  of $\Delta$ we expect, in first order, 
a linear behavior with $G_{trial}$ and this is what we obtain.  In general 
 a very broad range of gap parameters is covered,  which is the reason why the
 coordinate $G_{trial}$ can be considered equivalent to the gap parameter $\Delta$.
 The case of the coordinate $\mu$  is considered in panel b), where we represent the 
 corresponding gap parameter
as a function of  $\mu -\mu^N_{BCS}$. We find an oscillating behavior of $\Delta$ with
$\mu$ due to the  symmetry of the model. Notice that the scale of the y-axis depends on the 
particle number considered, see the figure caption.
For $\mu = kd$, i.e., at the
single particle energies $\epsilon_k$, we find maxima and for $\mu = k(d+1/2)$ minima. The period and amplitude
of the oscillations decrease with growing particle number because in this model $d \sim 1/N$.
For $N\geq 40$ we obtain superconducting solutions for all $\mu$ values, in particular for 
$\mu= \mu_{BCS}$, i.e., for the selfconsistent $BCS$ equation. For $N=20$, however, we observe
that at and around  $\mu = k(d+1/2)$ we do not obtain correlated wave functions. As mentioned
above this behavior is  in agreement with the fact that the selfconsistent BCS solution
does not have correlated solutions in this case. The situation is further
illustrated in Fig.~\ref{fig3} for
the $N=20$ case.  In the weak pairing regime we only find solutions for $\mu$ values corresponding to the
hatched regions around a given level. In the region around the level $k$, the number of particles of the $BCS$ w.f., 
i.e., the expectation value $\langle BCS(\mu) | \hat{N} | BCS(\mu) \rangle$, varies in a continuous way
from  $2(k-1)$ to $2k$. For example for the case of $N=20$, i.e. $k=10$, the $BCS$ w.f. around the level 10 have
average numbers of particles ranging from 18 to 20.  In general from these w.f.'s  it is always possible to project to
20 particles.  In the regions between the hatched regions no BCS solution
is found but only the Hartree-Fock (HF) one.  The numbers of particles are obviously 
 integer numbers, in the example displayed
these integers are $16, 18, 20$ and 22. To project to 20 particles from these HF w.f. is only 
possible for 20,  in the other cases the w.f. is
zero. This fact explains the curve corresponding to $N=20$ in Fig.~\ref{fig1}.  From all 
regions where no BCS solution
is found only  the one between $k=10$ and $k=11$, corresponding to a HF solution with 20 electrons with zero
condensation energy, survives. In Fig.~\ref{fig1} this region is represented by the straight line around $\mu= \mu_{BCS}$.
For $N\geq 40$ this is not the case and we always find $BCS$ solutions.  The hatched regions 
of Fig.~\ref{fig3} correspond
in this case to strong correlations and the white regions to weak ones.

Finally, in panel c) the gap parameters corresponding to 
the $\Delta N^2$ generator coordinate are plotted. The behavior is again linear, as for
the coordinate $G_{trial}$, but the range of the gap parameters involved in each wave function 
is the opposite one. In this case we obtain for small particle numbers a much larger range than
for  large particle numbers.
\begin{figure}[tbp]
\begin{center}
\includegraphics[width=65mm]{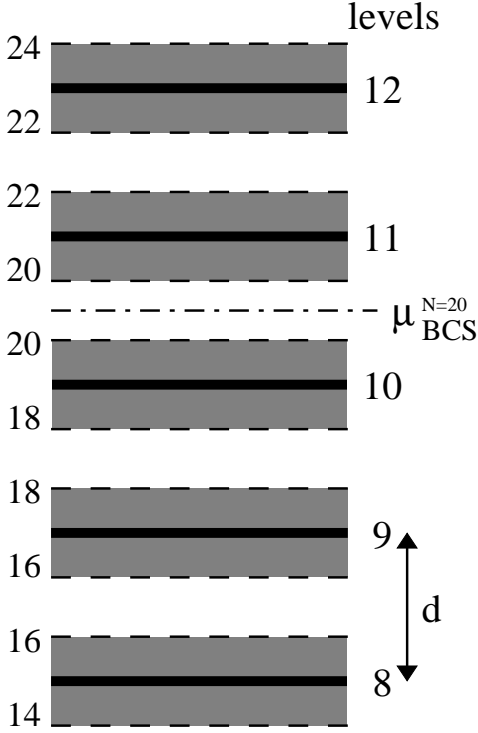}
\caption{ Sketch of the regions of weak and strong pairing for $N=20$.
The numbers on the right hand side correspond to the labels of the levels
while the ones on the left hand side to the average number of particles of the
BCS w.f. at the corresponding $\mu$.}
\label{fig3}
\end{center}
\end{figure}	 

\section{NUMERICAL SOLUTION OF THE HILL-WHEELER EQUATION}
\label{sect:num}
For our purposes solving the HW equation,   Eq.~\ref{HWEQ},  is equivalent to the
diagonalization of the Hamiltonian in the nonorthogonal basis of
the generator states $|BCS(\xi)\rangle_N$.
The usual procedure to deal with  this  equation \cite{RS.80} involves two diagonalizations.
In a first step,  the norm overlap $\mathcal{N}_{\xi \xi'}$ is diagonalized
\begin{equation}
\int d \xi' \mathcal{N}_{\xi \xi'} u_k(\xi')=n_k u_k(\xi),
\end{equation}
with the functions $u_k(\xi)$ forming a complete orthonormal set 
in the space of the weights $f(\xi)$.  Its eigenvalues
are never negative,  $n_k \geq 0$,  because the matrix $\mathcal{N}$
is definite positive. 
We shall keep the $u_k(\xi)$ with nonzero eigenvalues
corresponding to the linearly independent states.
In practice and due to numerical reasons we restrict  the $u_k(\xi)$ to those
with  eigenvalues larger than a tolerance $\varepsilon$.
For each of these functions there exist  states  $|k \rangle$,
\begin{equation}
| k \rangle = \frac{1}{\sqrt{n_k}} \int d \xi u_k(\xi) |BCS(\xi)\rangle_N,
\end{equation} 
called the \emph{natural states}, which span a collective subspace 
$\mathcal{H}_C$.  In a second step the  Hamiltonian $\hat{H}$ is diagonalized 
in this  space

\begin{equation}
\sum_{k'} \langle k |\hat{H}|k'\rangle g_{k'} = E g_k
\end{equation}
with
\begin{equation}
\langle k |\hat{H}|k'\rangle = \frac{1}{\sqrt{n_k n_{k'}}}\int \!\!\! \int
 d \xi d \xi' 
u_{k}^{\ast}(\xi) \mathcal{H}_{\xi \xi'} u_{k'}(\xi').
\end{equation}

The HW equations provide a set of wave functions, 
\begin{equation}
| \Psi^{\sigma}_N \rangle = \sum_{k,n_k \neq 0} g^{\sigma}_k |k\rangle,
\label{HWSOL}
\end{equation}
and energies $E_{\sigma}$ labeled by the index $\sigma$, the lowest one corresponding to
 the ground state and the others to excited 
states. In this work we are only interested in the ground state.
Taking into account Eq.~\ref{BCSANSA} and Eq.~\ref{HWSOL} one obtains
\begin{equation}
f(\xi) = \sum_k \frac{g_k}{\sqrt{n_k}} u_k(\xi).
\end{equation}
Since the wave functions $|BCS(\xi)\rangle_N$ are not orthogonal, the weights
$f(\xi)$ cannot be interpreted as the probability amplitude to find the
state $|BCS(\xi)\rangle_N$ in $| \Psi_N \rangle$. It can be shown, however,
that the functions
\begin{equation}
\mathcal{G} (\xi) = \sum_{k,n_k \neq 0} g_k u_k (\xi)
\label{collewf}
\end{equation}
are orthogonal and that they can be interpreted as probability amplitudes.

For numerical purposes all the expressions 
above involving integrals have to be  replaced
by sums discretizing the space $\xi$.  In this form one
 deals with matrix equations easier to handle.
The question that immediately arises is how to determine the optimal $\xi$-mesh
to be used in the calculation.  The border values of $\xi$ are determined 
by energy arguments since the probability of mixing high-lying states is very
small.  The $\xi$-coordinate intervals used in the calculations are given in Table~\ref{table2}.
The  calculations  depend furthermore on 
 the mesh step used in the discretization. This parameter is chosen as to optimize the calculations, i.e.,
we take the largest mesh that includes all states with  relevant information. This parameter is also related
to the required  accuracy. In the calculations performed we have not attempted to reproduce the exact
results up to an unusual accuracy.  
In Fig.~\ref{fig4} the convergence of the condensation energy, as a function of the
number of mesh points used in the calculations, is shown for different numbers of particles
and for the coordinates $G $, $\mu$ and $\Delta N^2$. 
We observe that the number of mesh points needed for convergence depends 
on the $\xi$-coordinate and on the number of particles. The coordinate 
$G_{trial}$, see top panel, provides the best convergence of the three calculations.
For large particle numbers  a very good 
convergence for relatively few mesh points is found.   For small numbers  of particle
one has to go to larger mesh points to find the plateau. For the $\mu$
coordinate the situation  is the reverse one, i.e., one finds
earlier  convergence for small numbers of particles. Finally, the situation for
$\Delta N^2$ is something in between the two former cases.  
 We find that to  reach convergence in energy 80 mesh points are sufficient for all coordinates.
 This is the number which we will use in all following numerical applications.

\begin{table}[bh]
\begin{center}
\caption{ Initial and final values of the generator coordinates used in the calculations.}
{\begin{tabular}{|c|c|c|c|c|c|c|c|c|c|c|} 
\hline
$$N$$ & $20$  & $40$ & $86$ &
$172$ & $400$ \\ \hline
$G_{i}(meV)$  &0.44& 0.18 & 0.07 & 0.03 & 0.01\\
$G_{f}(meV)$ & 0.80 & 0.30 & 0.17 & 0.06 & 0.04 
\\ 
$\mu_i(meV)$ & -3.00 & -2.00 & -2.00 & -1.00 & -1.00
\\
$\mu_f(meV)$ & 3.00 & 2.00 & 2.00 & 1.00 & 1.00
\\
$\Delta $N$^2_i$ & 0.00 & 0.00 & 0.00 & 0.00 & 0.00
\\
$\Delta $N$^2_f$ & 8.00 & 12.00 & 16.00 & 16.00 & 32.00
\\
\hline
\end{tabular}}
\label{table2}
\end{center}
\end{table}
\begin{figure}[tbp]
\begin{center}
\includegraphics[width=88mm]{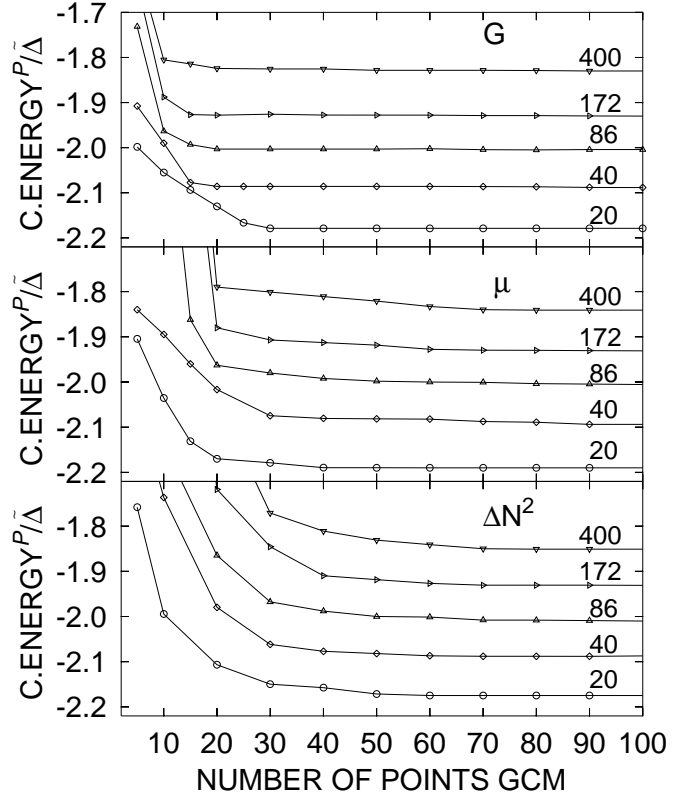}
\caption{ The condensation energy in units of $\tilde{\Delta}$ for the
three coordinates as a function of the number of mesh points used in the calculations.
 The energy scales correspond to the N=20 case, the other curves have been
 shifted in order to make the figure readable. The shifts are
  0.28,  0.58,  1.02 and  2.48 for 40, 86, 172 and 400 particles respectively.}
\label{fig4}
\end{center}
\end{figure}	 

A further check concerning the convergence is the number of natural states kept in the calculations.
 Since many of these states are linearly dependent some natural states 
$|k\rangle$ will have a vanishing norm and must be excluded. In the calculation only those natural states 
with a norm larger than a given tolerance $\varepsilon$ are kept. For a given tolerance we take as many  
states $|k\rangle$ as needed  to reach a good plateau.
Now we analyze the energy convergence as a function of the number of natural states
kept in the diagonalization of the HW equation or equivalently   of the tolerance of the
calculations.  This is shown in  Fig.~\ref{fig5}  for the three coordinates and for grains with different numbers of particles.
In the $\mu$ and $\Delta N^2$ coordinates we find that for tolerances smaller than $10^{-10}$ linear dependent states
are introduced in the calculations providing unrealistic energy values.  The interesting point is the nice plateau found
for larger tolerances. The tolerance of  $10^{-10}$  corresponds, typically, to around 15 linearly independent states.
 The coordinate $G$ is in this respect 
somewhat different.  One observes that for tolerances of up to  $10^{-15}$ one still gets linearly independent
states, which obviously  correspond to highly excited states that do not affect the energy of the ground state. This tolerance  typically 
amounts  to 20 linearly independent states. From this respect we conclude that if one is interested in excited states
the coordinate $G$ is more effective  than the other ones.

\begin{figure}[tbp]
\begin{center}
\includegraphics[width=88mm]{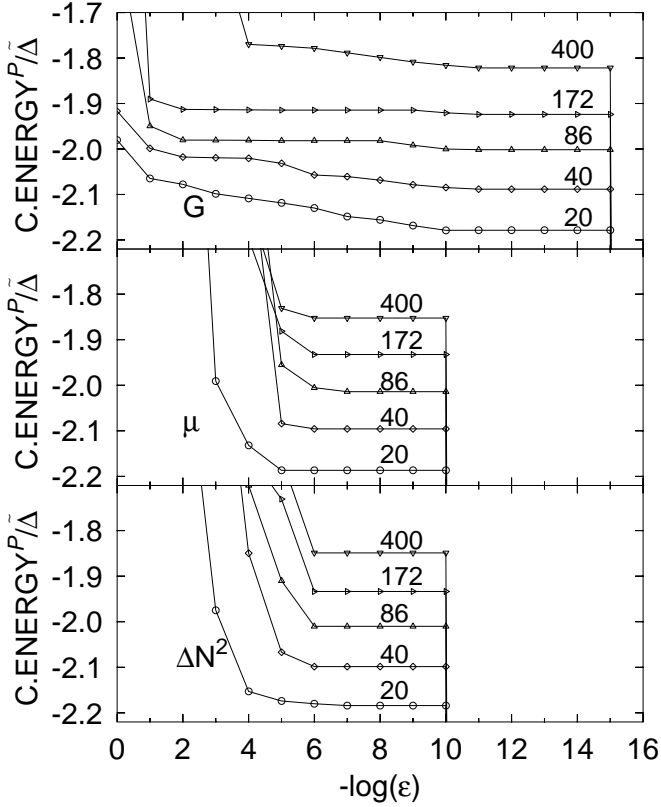}
\caption{Same as Fig.\ref{fig4} but as function of the tolerance $\varepsilon$.}
\label{fig5}
\end{center}
\end{figure}	 

The diagonalization of the HW equation in the case of the $\mu$ coordinate requires some comments. As mentioned
above  in the weak pairing regime, for $N=20$ for example, and for several $\mu$ intervals one does not find 
superconducting solutions.  This circumstance, as explained above, shows up as "missing points" in the CE curves. 
These discontinuities do not affect however the solution of the HW equations, since these points have norm zero and do no
mix with the other states.

\section{APPLICATION TO SUPERCONDUCTING GRAINS}  
\label{sect:app}
In this section we present a systematic study of properties of superconducting grains.  The results of the GCM calculation of
different quantities are compared  with the  PBCS approximation and the  exact  Richardson solution.
\begin{figure}[tbp]
\begin{center}
\includegraphics[scale=0.6]{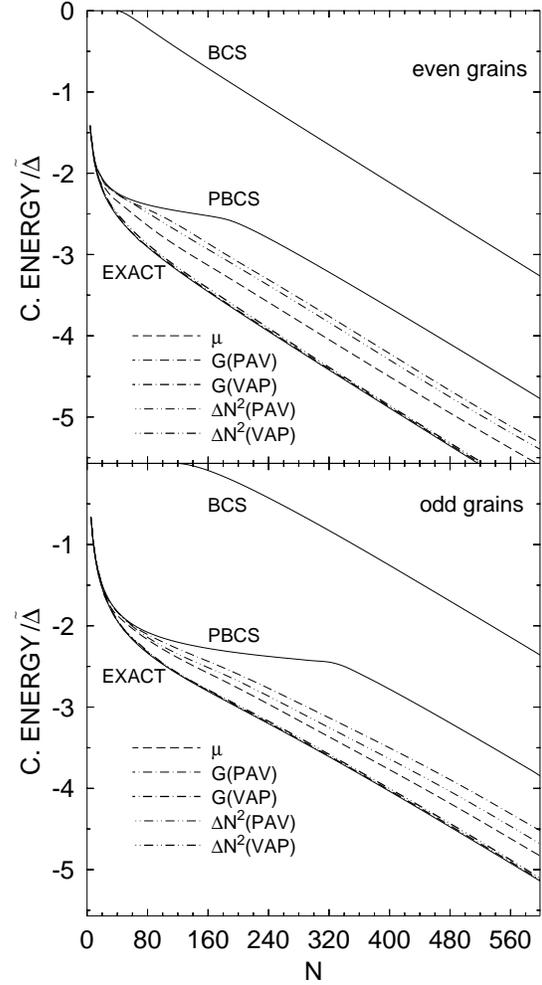}
\caption{Condensation energies versus the number of particles in different
approximations and the exact results. Upper (lower) panel for even (odd) systems.}
\label{fig6}
\end{center}
\end{figure}	 
\subsection{Ground State Condensation energies}
Condensation energies characterize the presence of pairing correlations. The crossover between
superconducting and fluctuation dominated regimes can be described through this quantity.

As in the former cases the condensation energy ${E}_{con}$ is defined as the difference between 
the total energy in the corresponding
approximation and the energy
of the uncorrelated Fermi sea. For example in the GCM approaches it is given by  
${E}_{con}= E_{\sigma=0} -E_{F}$, see Eq.~\ref{HWSOL} and below.
 This quantity is displayed in Fig.~\ref{fig6} for even grains (up to $600$ electrons)
 and  odd grains  (up to $601$ electrons)  as a function of the particle number $N$. In both
plots we give numerical results for the approximations
discussed above, BCS, PBCS and the GCMPAV and GCMVAP approaches.
The GCMPAV results are presented for the coordinates $G_{trial}$, $\mu$ and
$\Delta N^2$ and the GCMVAP for  $G_{trial}$ and $\Delta N^2$.  
The grand-canonical (BCS) calculation of ${E}_{con}$
predicts vanishing correlations in the few-electron regime in the even and odd systems.
  The PBCS condensation energies, on the other hand, though always negative
predict an unrealistic sharp crossover between the fluctuation dominated regime and the bulk which is more 
pronounced in odd grains. This artifact is not present neither in   the GCM approaches  nor  in the exact
 calculations.  The simpler GCMPAV  approaches already predict a smooth crossover for odd and even grains. 
 The more involved GCMVAP approaches not only predict a smooth crossover but their
predictions coincide with the exact results. Concerning the GCMPAV calculations we find that the
$\mu$ coordinate is the most effective of all of them followed by the $\Delta N^2$ one.  Paradoxically
the calculation with the $\mu$ coordinate is the simplest one from the numerical point of view.
 
 The reason why the $\mu$ coordinate is the most successful one is probably due to the fact that 
using this coordinate one has the right inertia parameter for the rotations in the gauge
space associated with the operator $\hat{N}$.
As a matter of fact  this was demonstrated by Peierls and Thouless \cite{PT.62}
in the  context of the translational invariance and a PAV approach by the double projection technique
(see  Eq.\ref{BCSANSA} above). In this work they show that the right inertia parameter of the 
collective motion associated with the linear momentum operator $\hat{P}$ ($\hat{N}$ in our case) 
is  obtained  when the GCM coordinates are the position ($\varphi$  in our case ) and the 
velocity ($\mu$ in our case). That means the dynamics associated with Eq.~\ref{HWEQ} has the right inertial
parameter.  In a VAP approach one always obtain the right mass parameter \cite{RS.80}.

\subsection{Pairing correlations.}
In a canonical ensemble the BCS order parameter is identically zero.
For this reason it is necessary to define  another quantity to characterize pair correlations 
in a state of fixed numbers of electrons.
We choose the pairing parameter used in ref. ~\cite{BD.98}
\begin{equation}
\Delta_b = G \sum_k C_k
\label{Deltab}
\end{equation}
where the subindex $b$ indicates the number parity of the grain. The quantities $C_k$'s are defined by
\begin{equation}
C_k^2 = \langle c_{k+}^{\dagger} c_{k+} c_{k-}^{\dagger} c_{k-}\rangle -
\langle c_{k+}^{\dagger} c_{k+}\rangle \langle c_{k-}^{\dagger}c_{k-}\rangle
\end{equation}
and form a set of  correlators which measure the fluctuations in the occupation numbers. The expectation
values $\langle \;\; \rangle$ are to be calculated with the wave functions of the corresponding approach using the formula developed in Appendices A and B.
In an uncorrelated or in a blocked state one has $C_k=0$.  In the grand-canonical case
the $C_k$'s reduce to $C_k = u_k v_k$ and $\Delta_b$ coincides with the usual
superconducting order parameter.

\begin{figure}[tpb]
\begin{center}
\includegraphics[scale=0.6]{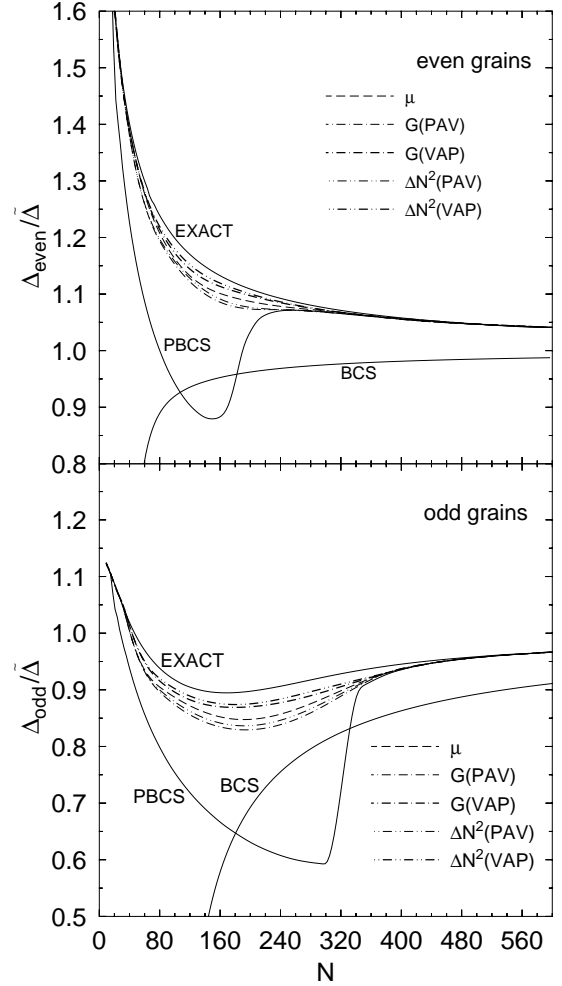}
\caption{The gap parameter $\Delta_b$ for the different calculations as a function of the number of electrons for even
and odd grains.  For particle numbers smaller than those shown in the BCS plot  the BCS gap parameter
goes sharply to zero.}
\label{fig7}
\end{center}
\end{figure}	 

In Fig.~\ref{fig7}  we show our results for the pairing parameter in units of $\tilde{\Delta}$ for even (upper panel) and odd (lower panel) systems respectively.  As we can see in both plots the sharp transition occurring
in the BCS and PBCS methods is absent in the GCM approaches as well as  in the exact solution. The peculiar behaviors of $\Delta_b$  in the exact and 
GCM approximations before and after the BCS breakdown  are related to  the change of a pairing delocalized in energy (weak pairing regime) 
to a localized one (strong pairing regime). The  rough decrease of  $\Delta_b$ with $N$ is connected to the special feature of the model, 
for which the constant $G$ of   Eq.~(\ref{Deltab}) is  inverse proportional to the
number of electrons. The fact that  $\Delta_b$ converges monotonically to the final value   $\tilde{\Delta}$ , 
in the even case from above and in the  odd one from below, is due to  the blocking effect.  In these plots we observe  again that the  GCMVAP approaches provide solutions closer to the exact one than the GCMPAV approaches.
\subsection{Collective wave functions.} 
We now look at the structure of the GCM states in the space  of the collective parameter $\xi$.
The collective weights $f(\xi)$  can not be interpreted as  probability amplitudes because the generating states 
$|BCS(\xi)\rangle_N$ are not, in general, orthogonal
to each other.
The amplitudes $\mathcal{G}(\xi)$ of  Eq.(\ref{collewf}), on the other hand, play the
role of "collective wave functions", they are orthogonal
and their modules squared have the meaning of a probability.

The quantities $|\mathcal{G}(\xi)|^2$ are plotted in Figs.~\ref{fig8}, \ref{fig9} and \ref{fig10}
as a function of the parameters $G$, $\mu$ and $\Delta N^2$, respectively.
 The behavior of $|\mathcal{G}(\xi)|^2$ as a function of $\xi$ indicates 
which are the most relevant components of the states $|\Psi_N(\xi)\rangle$ in terms of the parameter $\xi$.
 To guide the eye  we have also plotted in these figures the projected energy $E_N(\xi) $ of Eq.~(\ref{Eproj}).
 For simplicity we restrict our discussion to even systems.
In Fig.~\ref{fig8}  we represent these quantities for the coordinate $G_{trial}$ in the GCMPAV and GCMVAP
approaches.  Since the wave functions of both approaches do not differ qualitatively we shall discuss both
cases together.
The fact that the projected energies are lower in the PAV than in the VAP approach  for a given
$G_{trial}-G_c$ is due to the fact that $G_c$ is almost zero for all particle numbers in the VAP
approach while it varies considerably with the particle number for the PAV case, see  Table~\ref{table2}. We find
broad  potential energy curves for small particle numbers and  narrower ones with increasing $N$. Interestingly the 
potential energy curves  for the GCMPAV and GCMVAP approaches are rather different for small particle numbers and they 
become similar for the large ones.
Concerning the wave functions for $N=20$ we obtain  very broad distributions corresponding to a situation of weak pairing
 dominated by fluctuations in the order parameter $\Delta$, see also panel a) of Fig.~\ref{fig2}.  For $N\sim 40$
 we find that the wave functions are not that extended anymore but they still present a two peak distribution, wit the first
peak around the non-superconducting solution and the other around a superconducting one.  For larger particle
 numbers    ($N\sim 86, 172, 400$) a one peak distribution emerges with the width of the peak  getting smaller for increasing 
 particle number. Looking at Fig.~\ref{fig2}a we see that at large $N$ the distribution peaks around the wave function
with a gap very close to the bulk one.
\begin{figure}[tbp]
\begin{center}
\includegraphics[scale=.8]{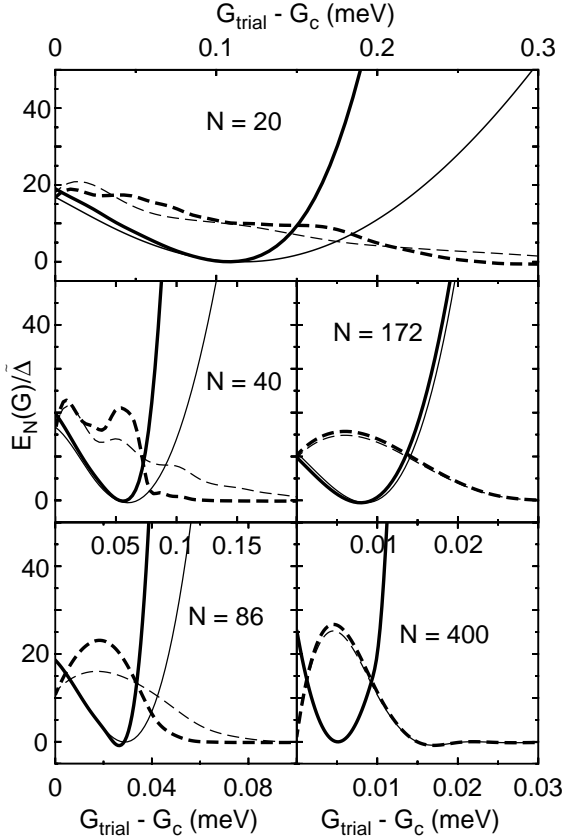}
\caption{The projected energies $E_N(G)$ versus $G$ in
the GCMPAV (thin continuous lines) and GCMVAP (thick continuous lines) approaches
for even systems.  The collective wave functions $|\mathcal{G}(G)|^2$ for
the GCMPAV (thin dashed lines) and the GCMVAP (thick dashed lines)
approaches in arbitrary units.  The vertical scale applies for $E_N(G)$, the minimum
of $E_N(G)$ in each approach has been set equal to zero. The top scale applies only for the top panel.}
\label{fig8}
\end{center}
\end{figure}	 

The results for the $\mu$ coordinate are presented in Fig.~\ref{fig9}. 
For $N=20$, in the weak pairing region,  we obtain a flat potential which shape
corresponds to the physics already discussed in relation with Fig.~\ref{fig3}. The
collective wave function, also according to  the discussion of Fig.~\ref{fig2}b and Fig.~\ref{fig3}, displays an
oscillating  behavior with maxima around $\mu=kd$ and minima (zeroes) around $\mu=k(d+1/2)$. The height of  the 
maxima decreases considerably for $k$ values different from 10 and 11, i.e., the collective wave function is mainly formed 
by the HF solution around the Fermi level and the N=20 components of the BCS solution of the levels above and
below the Fermi level.  For $N= 40$ the weak pairing regime persists and the wave function displays a structure
similar to the $N=20$ case  but with the strength much more concentrated owing to the fact that the level spacing
decreases with increasing number of particles. For $N=86$ the two peak structure is just a reminiscence of the
weak pairing situation and for $N=172 $ and 400 a one peak structure emerges indicating the strong pairing situation.
The potential energy curves get steeper with increasing N and the localization of the peak around $\mu^{N}_{BCS}$ 
sharpens in the same way. As one can see in Fig.~\ref{fig2}b the range of the $\Delta$ parameter covered by the
wave functions diminishes with increasing $N$.
 
\begin{figure}[tbp]
\begin{center}
\includegraphics[scale=0.8]{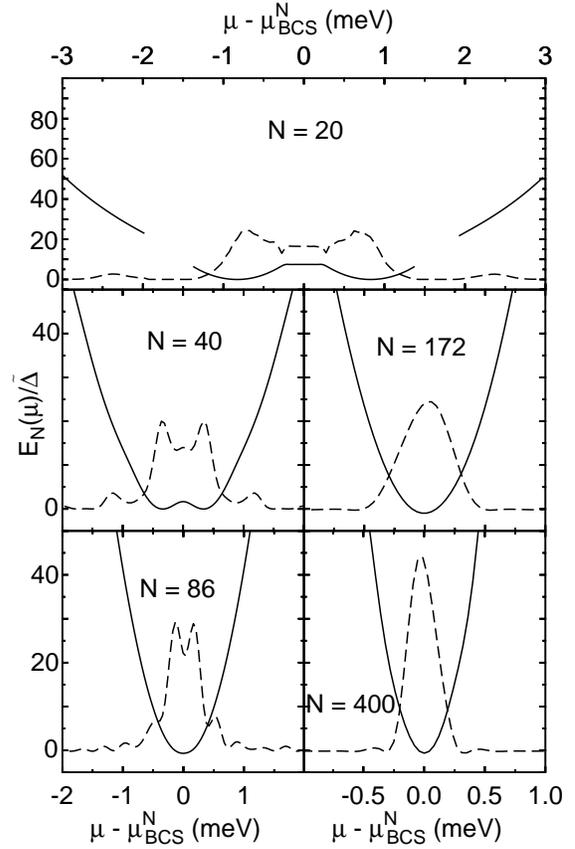} 
\caption{The projected energies $E_N(\mu)$   (continuous lines) and the collective wave functions $|\mathcal{G}(\mu)|^2$  ( dashed lines) versus $\mu$.
 The vertical scale applies for $E_N(\mu)$, the minimum of $E_N(\mu)$  has been set equal to zero. 
 $|\mathcal{G}(\mu)|^2$ is in arbitrary units.   The top scale applies only for the top panel.}
\label{fig9}
\end{center}
\end{figure}	 

Lastly we discuss the $\Delta N^2 $ coordinate in Fig.~\ref{fig10}.  The potential energy curves are easy to understand.
In the small particle number limit the BCS solution does not provide a superconducting solution and therefore
$\langle \Delta N^2 \rangle = 0$. On the other hand the BCS approximation provides the exact solution
in the bulk limit, i.e. there $\langle \Delta N^2 \rangle >>1$. Accordingly we expect minima in the projected energies
at small $\Delta N^2 $  for low N and at large $\Delta N^2 $ for large N. The potential energy curves get softer with growing
N because for increasing $\Delta N^2 $ it is energetically easier to change this value. As it should be the potential energy
curve for the GCMVAP approach lies below the GCMPAV one.  Concerning the collective wave functions, their behaviors correspond
to the shape of the potentials. For $N\le 86 $ there is a finite  probability of having an uncorrelated HF solution as a component 
of the collective wave functions and only for $N\ge 172$ we obtain Wigner-like functions with zero amplitude for the
HF component. The range of $\Delta$'s covered by the wave functions can be read from Fig.~\ref{fig2}c.

    Finally, we would like to mention that a very detailed comparison of our wave functions
and the exact ones has been made in ref. \cite{FE.03}. We find that the physical 
content of the GCMPAV wave functions and the exact ones is identical.

\begin{figure}[tbp]
\begin{center}
\includegraphics[scale=0.8]{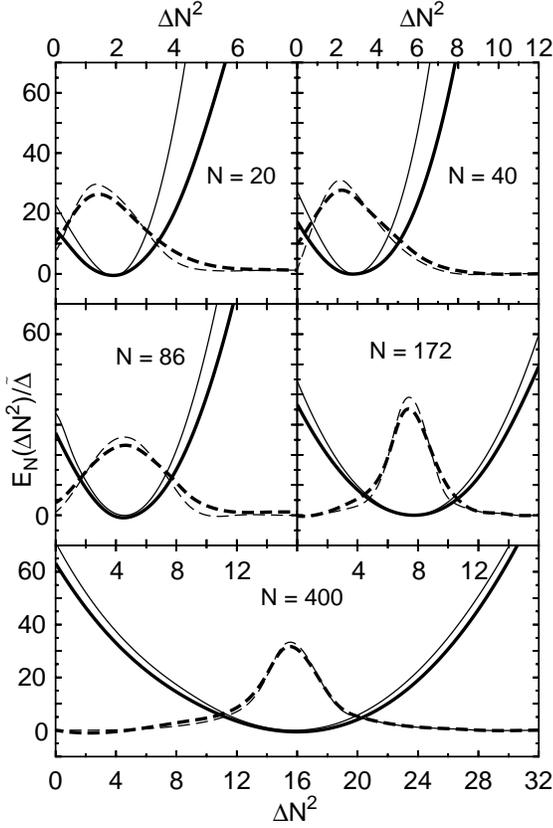}
\caption{The same as in figure \ref{fig8}  but for the parameter 
$\Delta N^2$.}
\label{fig10}
\end{center}
\end{figure}	 

\section{Conclusions}
 
 In this paper we have presented a detailed formulation of the particle number projected 
Generator Coordinate Method. 
   We have discussed two  different coordinates to generate wave functions for the variation after projection 
method and three for the projection after variation one.  The theory has been applied to study superconducting grains
with a pairing Hamiltonian. We have
shown that the GCMVAP calculations with both proposed coordinates reproduce the exact results
in the weak, crossover and bulk regimes. Concerning the GCMPAV calculations we find
that all three proposed coordinates, in spite of not being able to reproduce the exact results, describe qualitatively the 
correct physics washing out the phase transition found in the BCS and the PBCS approaches.  Concerning the degree of accuracy
we find  that the $\mu$ coordinate is the most effective of all the three followed by the $\Delta N^2$ one. 

  We think that these results are rather general and apply to many more complex  Hamiltonians  than the naive pairing one considered here.
Since the GCM Ansatz includes explicitly fluctuations in the wave function  it is very appropriate to deal with finite systems where 
phase transitions may take place.
  The method, contrary to other  approximations, applies equally well to systems with very few or very large particle  number.
   On the other hand the GCM Ansatz is very versatile to be adapted to other
physical situations by considering additional coordinates to the ones discussed in this work.
                   
This work has been supported in part by DGI, Ministerio de Ciencia 
y Tecnolog\'\i a, Spain, under Project FIS2004-06697. M.A.F. acknowledges a scholarship of the Programa
de Formacion del Profesorado Universitario (Ref. AP2002-0015)

\appendix                                                
\section{Peculiarities of the PBCS approximation}
In this appendix we discuss some numerical aspects of the solution of the 
equations used in this article.
\subsection{Odd particle number case}
\label{App:PBCSOdd}

The PBCS and HW methods can be extended to systems with an odd number 
of particles by blocking one of the available states. 
A system with an odd particle number is described by the state
\begin{eqnarray}
| BCS\rangle_{N+1}^l &=&  \int_0^{2\pi}  \frac{d\varphi}{2\pi} e^{iN\varphi}
c^{\dagger}_l \nonumber \\
&\times& \prod_{k\neq l} (e^{-i\varphi/2}u_k + e^{i\varphi/2}v_k
c^{\dagger}_{k+} c^{\dagger}_{k-})|-\rangle
\end{eqnarray}
with $N$ an even number and $l$ the blocked state.
 The residuum integral of Eq.~(\ref{resint}) now looks like
\begin{eqnarray}
^lR_m^{j_{1}, \cdots, j_{M}} 
&=& \frac{1}{2 \pi} \int_{0}^{2 \pi} d \varphi \, 
e^{-i(M - 2m)\varphi/2} \nonumber\\
&\times& \prod_{k \neq j_1, \cdots, j_M,l}^N (e^{-i\varphi/2}u_k^2 
+ e^{i\varphi/2} v_k^2),
\end{eqnarray}
in an obvious notation. As before all expectation values can easily be calculated  in terms of the residuum 
integrals, for example  the norm matrix element is given by
\begin{equation}
_{N+1}\langle BCS | BCS \rangle_{N+1} =  {^lR}_{0}^{0}, 
\label{NNormodd}
\end{equation}    
and the projected energy  by
\begin{eqnarray}
E_{N+1}^l  &=&   2 \sum_{j(\ne l)=1}^N \left( \epsilon_j - \frac{G}{2} \right) v_j^2 \frac{^lR_1^j}{^lR_0^0}  \nonumber \\
&-&  G \sum_{j,k\neq l}^N u_j v_j u_k  v_k \frac{^lR_1^{jk}}{^lR_0^0} + \epsilon_l. \nonumber
\label{Eprojodd}
\end{eqnarray}

The superindex is somewhat superfluous and it can be suppressed. Blocking  different states $l$ one obtains different excited
states.  The lowest of these energies corresponds to  the ground state of the system with an odd particle number.
The PBCS and HW variational equations obtained with theses states can be guessed without further calculations:
In all sums and products in the equations presented for even systems the term corresponding to the blocked state $l$ has to be excluded.
 In the same way one can write down states with 3,5,.. etc. blocked states.
 
\subsection{Numerical solution of the PBCS equations.}
\label{App:NumSol}

The set of coupled  non-linear equations (\ref{VAR_EQU}) is usually
solved by an iterative procedure. In order to speed up such procedure
it is convenient to introduce the  variable $\chi_k$ through the relation
\begin{equation}
u_k^2 = \frac{\chi_k}{1+\chi_k}, \,\,\,\,\, v_k^2 = \frac{1}{1+\chi_k},
\end{equation}
where the normalization condition \(u_k^2 + v_k^2 = 1\) has been taken into
account by construction. In terms of $\chi_k$ the variational equations
look like

\begin{equation}
\label{VAR_EQU_3}
\left(\hat{\epsilon}_k +  \Lambda_k \right)
\chi_k^{1/2} - \Delta_k
\left(\chi_k - 1\right) = 0.
\end{equation}

The additional transformation
\begin{equation}
\chi_k = \exp{\theta_k}
\end{equation}
allows to isolate the new variable in terms of  the
fields $\hat{\epsilon}_k, \Lambda_k$ and  $\Delta_k$
\begin{equation}
\label{VAR_EQU_FINAL}
\theta_k = 2 \sinh^{-1} \left( \frac{\hat{\epsilon}_k + \Lambda_k}
{2 \Delta_k}\right).
\end{equation}

The right hand side of this equation does not depend explicitly on $\theta_k$  because  the
fields $\hat{\epsilon}_k, \Lambda_k$ and $\Delta_k$ are independent of  $\theta_k$. This fact
is very useful to solve  Eq.~(\ref{VAR_EQU_FINAL})  by numerical iteration.  We start with a guess of $\theta_k$
(for instance the grand-canonical solution) and solve (\ref{VAR_EQU_FINAL}) until the convergence of the energy,
to a given tolerance,  has been reached. 

The variational equations (\ref{VAR_EQU_FINAL}) involve the computation of the 
residuum integrals $R_n^{\nu_1 \ldots \nu_M}$.  These   integrals can be calculated analytically
using the  existing  closed analytical expression \cite{TM.99}.  Their evaluation, however,  requires the addition of many terms making the 
whole computation a very time consuming approach.     As the integrals must be computed many times for different sets of $u_k,v_k$,
the numerical solution of  (\ref{VAR_EQU_FINAL}) in an efficient way requires the computation of the residuum integrals in a
fast and accurate way. For this purpose  we have implemented Fast Fourier Transform routines to evaluate the integrals reducing the number of
different integrals as much as possible to minimize the computational effort.  For this purpose the following two identities can be used.
The first one  was found by Dietrich et al \cite{DM.64}.  It can be shown that the residuum integrals satisfy  the following recursion relation
\begin{equation}
R_{m}^{j_1, \cdots, j_M} = u_k^2R_{m}^{j_1, \ldots, j_M,k} +
v_k^2R_{m+1}^{j_1, \ldots, j_M,k}
\end{equation}
The knowledge of two  residuum integrals allows to calculate a different one. This   reduces the number of numerical integrations by one third. 
A second, more powerful  relation was found by Ma and Rasmussen \cite{MR.77},
\begin{eqnarray}
R_m^{j_1, \ldots, j_M} &=& \delta_{m,M}R_0^0 \!\!\!\!\!\!\!\! 
\prod_{j=j_1,\ldots, j_M}
\frac{1}{v_j^2}   
+ (-)^m \!\!\!\!\!\!\! \sum_{j=j_1, \ldots, j_M} \!\!\!\!\!\! 
v_j^{2(M-m-1)}u_j^{2m} \nonumber \\
&\times& \left( \prod_{k=j_1,\ldots, j_M(\neq j)}\frac{1}{v_j^2 - v_k^2}\right)
R_0^j .
\end{eqnarray}

This  formula allows to calculate {\em all} residuum integrals if the integrals  $R_0^0$ and $R_0^j$ are known.  This relation  reduces to  $N+1$
the overall number of numerical integrations for a given set of $v_j$'s and  $u_j$'s.

In the PBCS one only needs Ma's relation to calculate three terms: $R_1^{jk}$,$R_2^{jkl}-R_1^{jkl}$ and $R_2^{jk}-R_1^{jk}$.  If $R_0^0$
and $R_0^j$ are known, $R_1^j$ can be obtained by
Dietrich's recursion relation.  First we consider $R_m^{jk}$
with $m=1$ or $m=2$.  Ma and Rasmussen's formula reduces to
\begin{equation} 
R_{m}^{jk} = \delta_{m,2}\frac{R_0}{v_j^2v_k^2} + (-)^m
\frac{\zeta_j^m v_j^2 R_0^j - \zeta_k^m v_k^2 R_0^k}{v_j^2 - v_k^2}
\end{equation}  
where we have used the  identity 
$\zeta_j = \frac{1}{v_j^2} -1$. The difference $R_2^{jk} - R_1^{jk}$ can be written in a simplified way as follows
\begin{equation}
R_2^{jk} - R_1^{jk} = \frac{R_0}{v_j^2v_k^2} - \frac{R_0^j(v_j^2 -1)}
{v_j^2(v_j^2-v_k^2)} + \frac{R_0^k(v_k^2-1)}{v_k^2(v_j^2-v_k^2)}
\end{equation}

The calculation of $R_2^{jkl} - R_1^{jkl}$ is a bit more complicated,  the  result is
\begin{eqnarray}
R_2^{jkl}-R_1^{jkl} &=& \frac{u_j^4 + u_j^2v_j^2R_0^j}{(v_j^2-v_k^2)(v_j^2-v_l^2)}+
\nonumber
\frac{u_k^4+u_k^2v_k^2R_0^k}{(v_k^2-v_j^2)(v_k^2-v_l^2)} + \\
&+&\frac{u_l^4+u_l^2v_l^2R_0^l}{(v_l^2-v_j^2)(v_l^2-v_k^2)}.
\end{eqnarray}
Since the indices of the residuum integrals can be permuted,  $R_2^{jkl} - R_1^{jkl}$ can be expressed for all
possible combinations of $v_j,v_k,v_l$ by the  equations above.

\section{The generalized residuum integrals} 
\label{app:HamOver}
The Hamiltonian overlap $\mathcal{H}_{\xi \xi^{'}}$, Eq.~(\ref{hamove}),  can be calculated using the generalized Dietrich's recursion relation,
\begin{eqnarray} 
R_{m}^{j_{1}\cdots j_{M}}(\xi,\xi') &=& u_k(\xi)u_k(\xi') R_{m}^{j_{1}
\cdots j_{M},k}
(\xi,\xi') + \nonumber \\
&+& v_k(\xi)v_k(\xi')R_{m+1}^{j_{1}\cdots j_{M},k}(\xi,\xi')
\end{eqnarray}

The residuum integrals $R_1^{i},R_1^{ij}$ needed  to calculate  $\mathcal{H}_{\xi \xi^{'}}$ can be written in terms of $R_0^j$ and
$R_0^{jk}$ using Dietrich's relation twice 
\begin{equation}
\label{Di1}
R_1^{k}(\xi,\xi') = \frac{R_0^0(\xi,\xi') - 
u_k(\xi)u_k(\xi')R_0^{k}(\xi,\xi')}{v_k(\xi)v_k(\xi')}
\end{equation}
\begin{equation}
\label{Di2}
R_1^{jk}(\xi,\xi') = \frac{R_0^{j}(\xi,\xi') - u_k(\xi)u_k(\xi')
R_0^{jk}(\xi,\xi')}{v_k(\xi)v_k(\xi')}.
\end{equation}
The calculation of the Hamiltonian overlap using this method is rather slow.  In order to   speed up the  evaluation of the residuum integrals,   we have written 
the Hamiltonian overlap $\mathcal{H}_{\xi \xi^{'}}$  in the form
\begin{equation}
\label{hahap}
\mathcal{H}_{\xi \xi^{'}} = \frac{1}{2\pi}\int_{0}^{2\pi}\!\! d 
\varphi \, [F_1^{\varphi}(\xi,\xi^{'}) - F_2^{\varphi}(\xi,\xi^{'}) ]
\end{equation}
where

\begin{eqnarray}
&&F_1^{\varphi}(\xi,\xi^{'}) = 2 \sum_j (\epsilon_j - G/2)v_j(\xi)v_j(\xi^{'})
e^{i \varphi/2} \nonumber \\
&\times& \prod_{k\neq j}(e^{-i\varphi/2}u_k(\xi)u_k(\xi^{'})+
e^{i\varphi/2}v_k(\xi)v_k(\xi^{'}))
\end{eqnarray}

\begin{eqnarray}
&&F_2^{\varphi}(\xi,\xi^{'}) = G \sum_{ij} 
u_i(\xi^{'}) v_i(\xi) u_j(\xi) v_j(\xi^{'})  \nonumber\\
&\times& \prod_{k\neq ij}
(e^{-i\varphi/2}u_k(\xi)u_k(\xi') + e^{i\varphi/2}v_k(\xi)v_k(\xi')).
\end{eqnarray}
In the actual calculations  we used these expressions to evaluate $\mathcal{H}_{\xi \xi^{'}}$ instead of
Eqs. (\ref{Di1})(\ref{Di2}), the  integrals \ref{hahap} are evaluated by fast Fourier routines. 


\end{document}